# Tailoring Plasmonic Metamaterials for DNA Molecular Logic Gates

Jun Zhang[1,#], Cuong Cao[1,#], Xinlong Xu[2], Chihao Liow[3], Shuzhou Li[3], Ping-Heng Tan[4], and Qihua Xiong[1,5,*]

[1]Division of Physics and Applied Physics, School of Physical and Mathematical Sciences, Nanyang Technological University, Singapore 637371

[2]Nanobiophotonic Center, State Key Lab Incubation Base of Photoelectric Technology and Functional Materials, Institute of Photonics & Photon-Technology, Northwest University, Xi'an 710069, China

[3]Division of Materials Science, School of Materials Science and Engineering, Nanyang Technological University, Singapore 639798

[4]State Key Laboratory of Superlattices and Microstructures, Institute of Semiconductors, Chinese Academy of Sciences, Beijing 100083, China

[5]NOVITAS, Nanoelectronics Centre of Excellence, School of Electrical and Electronic Engineering, Nanyang Technological University, Singapore, 639798

[*]To whom correspondence should be addressed. Email address: Qihua@ntu.edu.sg.

[#]These authors contributed equally to this work.

**Structural and functional information encoded in DNA combined with unique properties of nanomaterials could be of use for the construction of novel biocomputational circuits and intelligent biomedical nanodevices. However, at present their practical applications are still limited by either low reproducibility of fabrication, modest sensitivity, or complicated handling procedures. Here, we demonstrate the construction of label-free and switchable molecular logic gates that use specific conformation modulation of a guanine- and thymine-rich DNA, while the optical readout is enabled by the tunable alphabetical metamaterials, which serve as a substrate for surface enhanced Raman spectroscopy (MetaSERS). By computational and experimental investigations, we present a comprehensive solution to tailor the plasmonic responses of MetaSERS with respect to the metamaterial geometry, excitation energy, and polarization. Our tunable MetaSERS-based DNA logic is simple to operate, highly reproducible, and can be stimulated by ultra-low concentration of the external inputs, enabling an extremely sensitive detection of mercury ions.**

Since the first introduction by Adleman in 1994[1], the DNA logic gates have been considered as the future of computation technology where their small size is a distinct advantage over the conventional top-down semiconductor technology[1-3]. They also exhibit important applications in life sciences as smart sensing and diagnostic platforms owing to the unique properties of DNA such as self-assembly, specific recognition and conformation modulation upon exposing to external stimuli (*i.e.* metallic ions, proteins)[4-6]. Along this line, DNA-based systems (*e.g.* DNAzymes[5, 6], molecular beacons[3, 7], Guanine-rich oligonucleotides (G-quadruplexes)[8], aptamers[9]) have been devised on different nanometer scale carriers such as graphene[10], graphene oxide[9, 10], solid-state nanochannels[11], quantum dots[12], and gold nanodisc arrays[13] for various logic gate operations and biosensing applications. Those DNA logic operations mostly rely on fluorescence and enzyme cascades to generate "ON" or "OFF" output signals which involve complex handling and analysis procedures, thus restricting the performance and applications of the sophisticated logic devices. In addition, it still remains very challenging to realize a label-free and switchable DNA logic gate-based biosensing platform that can selectively respond to extremely low concentration of the chemical and biological stimuli.

To extend new dimensions towards this rapidly growing area, our work is motivated by the recent emergence of plasmonic metamaterials capable of providing high electromagnetic enhancement (hot-spots) for surface enhanced Raman scattering (hereafter called MetaSERS)[14-16]. The most prevailing SERS sensors are based upon chemically synthesized colloidal nanoparticles[17, 18], while the disadvantage is the poor reproducibility. Recent advances have employed a variety of top-down fabrication techniques which enable large-scale and reproducible patterns for SERS substrates ranging from bow-tie nanoantennae[19, 20] to asymmetric Fano resonance structures[21, 22]. In contrast to the above structures, plasmonic metamaterials have recently been demonstrated to offer effective ways to tailor the concentration of light to form desired hot-spots by controlling the size and shape of plasmonic structures[23]. By properly designing the micro- or nano-scaled metallic sub-wavelength structures, *i.e.* "split ring resonators (SRRs)", one can tune the operating frequency of metamaterials from microwave[24] to visible regime[14, 25]. However, most of metamaterials-based biosensing focuses on the localized surface plasmon resonance (LSPR) shifts induced by absorbed molecules[26, 27], in which the shift depends on the effective refractive index of the target molecules thus exhibiting no chemical fingerprints. Recently, Fano-resonant asymmetric metamaterials have been introduced to demonstrate the ultrasensitive sensing and identification of molecular monolayer by tuning the resonant peak towards (away from) protein's vibrational fingerprint and monitoring the infrared reflectance spectra[22]. Here we demonstrate that tuning the Vis-NIR alphabetical metamaterials modulates the optical response and hot-spots leading to an ultrasensitive SERS detection. We subsequently describe INHIBIT and OR logic gate operations based upon the metallophilic properties of a guanine- and thymine-rich oligonucleotide sequence to $K^+$ or $Hg^{2+}$ ions, which can specifically trigger or interrupt the formation of Hoogsteen hydrogen bonding that could be monitored by means of MetaSERS with high sensitivity and selectivity. Significantly different from many other fluorescence-based or DNAzyme-based logic gate operations which involve complex handling and analysis procedures, the MetaSERS is a direct measurement and can be implemented without the need of any labeling fluorescent dyes or enzymatic activities. Moreover, the molecular logic enables the ultrasensitive detection of mercury ions at a concentration as low as $2\times10^{-4}$ ppb, which is at least 3 orders of magnitude improvement compared to literatures[7, 12, 28-31].

**Resonance modes identification of alphabetical metametrials.** Fig. 1 shows a series of alphabetical metamaterial structures (Fig. 1a), their optical responses (Fig. 1b), and mode identification (Fig. 1c). The fabrication details can be found in Methods. In order to obtain the tunable optical response from Vis to NIR range, we shrink the whole unit cell along with the dimension of resonators from 100% to 37.5%, which reduces the corresponding bar width ($w$) from $w$ = 80 to 30 nm (Methods). Fig. 1b shows the transmission spectra for various alphabetical metamaterials with different widths. The solid curves are spectra taken under $P_x$ polarization configurations, while the dashed curves are obtained from $P_y$ polarization configurations. Each valley in the transmission spectra represents one resonance mode of metamaterials. The induced electric dipoles due to the alternating field of light lead to two kinds of resonance modes: one is electric mode arising from oscillation of the electric dipoles, while the other is the magnetic mode due to the circular currents induced by head-to-end electric dipoles configuration[32]. For all five shapes, the weak short-wavelength resonances around 550 nm exhibit independence on the width, which are actually attributed as the dipole excitation along the width of arms[25]. The other resonance modes exhibit a systematic blue shift as the width decreases, labeled with color lines as guide to eye. This can be explained as follows. For electric modes, resonance frequency is proportional to the coupling strength of electric dipoles[32]. With a decreasing of the metamolecule width, the coupling strength and thus the resonance frequency increase. The magnetic mode can be seen as the analogy of inductor-capacitor circuit (LC) resonance[33]. With the decrease of the metamolecule size and bar width, the capacitance $C$ and inductance $L$ decrease and the resonance frequency increases. Fig. 1c shows the schematic diagram of the current distribution at the resonance wavelength based on the simulation of discrete dipole approximation (DDA)[34, 35] and dipole-dipole coupling theory[32]. The black arrows in Fig. 1c indicate the current direction. The detailed DDA simulation results of 30 nm width are shown in Fig. S2. The simulated mode numbers, polarization and relative energy are in good agreement with experimental results within 10% deviation (supplementary information).

Now we identify the physical origin of all the labeled modes. The U shape resonator is a typical SRR and has been extensively studied due to its negative refractive properties[25, 33]. It has three resonance modes from high to low energy as shown in Fig. 1c: the higher order magnetic resonance ($M_{h\text{-order}}$ modes, highlighted by the solid blue line in Fig. 1b), fundamental electric resonance ($E(y)$ modes, highlighted by dash green line in Fig. 1b, y represents the polarization direction of incident light), and fundamental magnetic modes ($M(z)$, highlighted by solid red line in Fig. 1b, z represents the direction of magnetic dipole moment). Here, we identify the highest frequency mode as the higher order magnetic resonance[25] rather than electric plasmon mode[36] because it has a partial circular current, leading to a magnetic-dipole moment as shown in Fig. 1c and Fig. S2. Based on the simulation shown in Fig. S2, we conclude that $M_{h\text{-order}}$ and $M(z)$ modes can only be excited at the $P_x$ polarization configuration (electric field vector of incident laser is along the *x*-direction) while $E(y)$ mode can only be probed at the $P_y$ polarization configuration.

The analysis of U-shaped resonator can be applied to interpret the other shapes. For instance, the Y-shaped resonator can be regarded as two connected U-shaped resonators rotated by a 90°. The modes in the isolated U-shape will couple together and lead to new modes in the composite shapes. The solid blue line highlights the size dependent evolution of the coupled higher order magnetic mode $M_{h\text{-order}}$, which is induced by $P_x$ polarized light and shows more complex current distribution as shown in Fig. 1c and Fig. S3b. The $P_x$ and $P_y$ polarization of incident light induce a couple of degenerated electric mode $E_{sym}(x)$ and $E_{sym}(y)$ corresponding to the valleys highlighted by the green solid and dashed lines in Fig. 1b, respectively. This is supported by our simulation (Fig. S3). However, experimentally we have observed the mode splitting as the Y-shaped resonator size increases, which are highlighted by the green solid and dashed lines in Fig. 1b. We speculate that as the size increases the degeneracy is lifted due to non-ideal symmetry of the two U-shaped resonators. For magnetic modes, the fundamental magnetic dipole modes in each U-shaped resonator couple together, resulting in the two degenerate symmetrically coupled magnetic dipoles modes ($M_{sym,x}(z)$ and $M_{sym,y}(z)$) and one asymmetric mode $M_{asym}(z)$. As there is a phase retardation between two 90°-rotated U shapes, the coupling between two magnetic dipoles leads to the spectral splitting of resonance[35]. Based on the dipole-dipole coupling theory[32], the north and south poles of the two neighboring magnetic dipoles repel each other in the symmetric mode, leading to the longer wavelength $M_{asym,y}(z)$ mode as highlighted by the dashed red line in Fig. 1b under a $P_y$ excitation, which is further red shift beyond the range of our spectrometer in resonator sized larger than ~ 60 nm. In a $P_x$ polarization excitation, only $M_{sym,x}(z)$ mode is observed as highlighted by solid pink line in Fig. 1b. With a similar argument, the spectra of coupled magnetic modes ($M_{sym,x}(z)$ and $M_{sym,y}(z)$) also split into two with the increasing of the size. Similar analysis can be applied to S, H, U-bar, and V shaped resonators (more details refer to Supplementary Information).

**Maximization of the SERS enhancement by tuning the laser wavelength.** Alphabetical metamaterials operated in visible-IR exhibit abundant electric and magnetic dipole modes, and their coupling effect gives further degree of freedom to tune the plasmonic resonance to optimize the SERS effect. The versatile tunability enables the maximization of the strength of local electromagnetic field hot-spots, which dominate the electromagnetic enhancement in SERS effect. We first show how to obtain the highest SERS signal by tuning the laser wavelength for H-shaped metamaterials as an example. Fig. 2a and 2b display the typical SERS spectra of a monolayer 2-naphthalenethiol bound to H40 sample excited by a tunable laser with $P_x$ and $P_y$ polarizations, respectively. Considering that laser wavelength (660-840 nm) is far from the first electronic transition (~242 nm) of 2-naphthalenethiol[37], we propose that the enhancement of SERS signal is entirely contributed by the electromagnetic enhancement. We chose the Raman peak around 1,380 cm$^{-1}$ originated from the ring-ring stretching mode to investigate the resonant SERS profile depending on the laser wavelength[37]. Its integrated area intensity is plotted in Fig. 2c and 2d along with comparison data from H50 sample. By tuning the laser wavelength and polarization to match the corresponding resonant modes in metamaterials, the enhancement was obtained about 20 times as compared to that of the off-resonance case. Based on the Mie scattering theory[38], the electromagnetic enhancement factor (*EF*) is the product of incident light and scattered light enhancement, *i.e.* $EF_{total} = EF(\lambda_{laser}) \times EF(\lambda_{scatt})$, where $EF(\lambda)$ has the same dispersion relationship with the extinction spectra of metameterials. In the case of $P_x$ polarization, the $M_{h\text{-order}}$ (~730 nm for H40, and ~820 nm for H50) and $M_{asym}(y)$ (~1561nm for H40, and ~1898 nm for H50) were excited (as show in Fig. 1b). Because the $M_{asym}(z)$ mode is far from the excitation laser and the corresponding stokes shift (1,380 cm$^{-1}$), the stronger signal at higher energy excitation shown in Fig. 2c is due to the $M_{h\text{-order}}$ resonance. While in the $P_y$ excitation, two other modes of $E_{sym}(y)$ (~780 nm for H40, and ~900 nm for H50) and $E_{asym}(y)$ (~860 nm for H40, and ~1150 nm for H50) are excited. As shown in Fig. 2d, two resonant peaks are observed in the H40 sample. One resonant peak is very sharp at around ~790 nm with a FWHM of ~10 nm. We propose that it is a double resonant process, in which the incident laser at ~785 nm resonates with $E_{sym}(y)$ while the scattered light at ~887 nm (~1,380 cm$^{-1}$) also within the resonance of the $E_{asym}(y)$ mode. The other peak around 710 nm is much broader because only the scattered light around 787 nm (1,380 cm$^{-1}$) can resonate with the $E_{sym}(y)$ mode. For H50 sample, the double-resonance is relaxed, which results in one broad resonant peak around ~700 nm because its $E_{asym}(y)$ mode is too broad and far from the scattered light. Accordingly, we have calculated the resonant profiles of enhancement by multiplying the simulated $|E/E_0|^2$ of the extinction spectra at the incident laser wavelength and the scattered light wavelength for the 1,380 cm$^{-1}$ mode, for both H40 and H50 samples. The results are shown in Fig. 2e and 2f. As can be seen, the simulated results are qualitatively in good agreement with the experimental results. The difference may come from the relative deviation of resonance peak position between experiments and theory.

**Maximization of the SERS enhancement by tuning the shape and the size of metamaterials.** For a given laser wavelength, for instance 785 nm, the enhancement can be maximized by tuning the size and shape of the alphabetical metamaterials. In this context, the chosen patterns are U, V, H, S and Y shaped metamaterials with bar-width from 30 to 50 nm, also functionalized with a layer of 2-naphthenethiol molecules. Fig. 3a shows a 2D plot of the SERS spectra under two different polarizations. Bar charts in Fig. 3b statistically summarize the simulated and experimental intensities for all different shapes and sizes. It becomes pronounced that the U40, V30, H40, S40 and Y30 samples show much stronger intensities than their other counterparts under a certain polarization. This suggests the versatility and tunability of metamaterials for a known laser excitation in order to gain the highest SERS effect. The explanation can be supported by the simulation of the local electric field contour distribution (or saying the hot-spots) as shown in Fig. 3c. One can see that both experimental data and simulation agree very well with each other. For the 785 nm laser, the simulated highest and average enhancement factors are ~3×10$^6$ (V30 for $P_y$ polarization) and ~25,900 (U40 for $P_x$ polarization). By comparing the Raman spectra of 2-naphthalenethiol powder and the SERS spectra of a covalently self-assembled monolayer of 2-naphthalenethiol, we could also estimate the average enhancement factors (averaging all area of metamaterials pattern) of experimental SERS spectra which are ~$10^6$ to ~$10^8$, respectively (See supplementary information and Fig. S4). These enhancement values are strong enough for detecting a few molecules located within proximity of the hot-spots.

**Construction of DNA logic operations.** We understand from previous discussions that the U40 metamaterials provide the highest electromagnetic enhancement as excited by a 785 nm laser. Such strong enhancement allows the alphabetical metamaterials to be exploited as DNA molecular logic circuits based upon SERS effect. As shown in Fig. 4a, the principle of the logic operations is based on the sequentially coordinating effects of $Hg^{2+}$ and $K^+$ on the conformational modulation of a short guanine (G)- and thymine (T)-rich oligonucleotide sequence ($(GTT)_4TG(TGG)_4$). In the presence of $K^+$ cations, G-rich oligonucleotides are known to fold into a specific and stable three-dimensional shape, namely G-quadruplex where four G can self-assemble to form a distinct Hoogsteen hydrogen-bonded square (*i.e.* G-tetrad or G-quartet) *via* C8=N7-H2[39]. The Hoogsteen hydrogen bonding results a sharp and strong peak centered at ~1,485 cm$^{-1}$ as measured by Raman scattering spectroscopy[40, 41]. On the other hand, $Hg^{2+}$ cations have been demonstrated to bridge specifically with two thymines by labile covalent bonds *via* N–$Hg^{2+}$ to form a hairpin T–$Hg^{2+}$–T complex with a binding constant of ~$8.9\times10^{17}$ M$^{-1}$,[7] which is much higher than that of $K^+$ cations stacking with the quadruplex structure (~$5\times10^6$ M$^{-1}$)[42]. Therefore, it provides a rationale for a DNA-based detection of $Hg^{2+}$ in which the formation of T–$Hg^{2+}$–T complex in the presence of $Hg^{2+}$ will inhibit the formation of G-quadruplex structure that in turn leads to the diminishment of the diagnostic Hoogsteen hydrogen bonding at ~1,485 cm$^{-1}$. In addition, it has been previously reported that the formation constant of $Hg^{2+}$ ions and iodide ($I^-$) is as high as $5.6\times10^{29}$,[43] therefore the introduction of $I^-$ could competitively disrupt the T–$Hg^{2+}$–T bonding and lead to the reversible formation of the Hoogsteen band under the presence of $K^+$. This will generate reversibly the combinational INHIBIT and OR logic gates schematically represented in Fig. 4b.

Fig. 4c and 4d depicts the MetaSERS-based INHIBIT logic gate operations upon interaction of the GT-rich oligonucleotide with the inputs $Hg^{2+}$ and $K^+$. The presence and absence of $Hg^{2+}$ or $K^+$ are defined as 1 and 0, respectively. The intensity of Raman band at ~1,485 $cm^{-1}$ normalized by its full width at half maximum (FWHM) is represented for the output 1 or 0. The combinations of four possible inputs are listed in the truth table (Fig. 4e). In the absence of inputs (0, 0) or with the $Hg^{2+}$ alone (1, 0), the GT-rich oligonucleotide is respectively in its unfolded state or in complexed form with T–$Hg^{2+}$–T, and therefore no strong Hoogsteen band is observed. The intensity of the Hoogsteen hydrogen band is significantly increased when the $K^+$ ions are introduced (0, 1), giving rise to the output of 1. However, the band at ~1,485 $cm^{-1}$ is completely diminished in the presence of both inputs (1, 1), meaning that the coordination of $Hg^{2+}$ in the T–$Hg^{2+}$–T complex inhibits the formation of the Hoogsteen hydrogen bonding. The assignment of a few other strong peaks in the spectra will be discussed in detail in another work[44].

As a means to evaluate the reversibility of the Hoogsteen hydrogen bonding, $I^-$ ions are subsequently introduced to the logic operation (Fig. 4f-h). In this case, the output of the INHIBIT logic gate (output 1) is used as one of the inputs. Fig. 4f and 4g show the SERS spectra and the normalized Raman intensities at ~1,485 $cm^{-1}$ for monitoring the reformation of the Hoogsteen band. It should be noted that $K^+$ has been already introduced in the buffer solution, thus G-quadruplex formation will be generated as long as the free G-rich DNA is present. $I^-$ ions strongly bind with $Hg^{2+}$ to break the bridge between thymine and $Hg^{2+}$ and to liberate the GT-rich oligonucleotide in such a way that the G-quadruplex could be reformed by stacking with $K^+$.

Therefore, in the absence of inputs (0, 0) the T–$Hg^{2+}$–T complex could not be disrupted, resulting in the output 2 of 0. However, the output 2 is true if either the output 1 or $I^-$ ions is true ((0, 1), (1, 0), or (1, 1)), leading to an OR logic operation as shown in the truth table (Fig. 4h).

**MetaSERS-based DNA logic gate for ultrasensitive detection of mercury ions.** Bivalent mercury ions $Hg^{2+}$ are the most stable inorganic forms of mercury contaminant in environment and food products and are responsible for a number of life-long fatal effects in human health such as kidney damage, brain damage, and other chronic diseases[45-47]. According to the United States Environmental Protection Agency (EPA), the maximum amount of mercury should be lower than ppb and ppm levels for drinking water and food products, respectively[45, 46]. However, because mercury has a strong bioaccumulation effect through the food chain[47], therefore there exists a great demand and also a significant challenge for development of a method that allows facilely monitoring the concentration of mercury below the defined exposure limit level. The principle of molecular logic gates discussed in Fig. 4 presents a rationale for ultrahigh sensitive detection of $Hg^{2+}$ ions: the trace amount of $Hg^{2+}$ ions bind to the GT-rich oligonucleotides to form hairpin structures thus strongly inhibiting the formation of the quadruplex structures. As a result, the 1,485 $cm^{-1}$ Raman fingerprint of the Hoogsteen hydrogen bonding diminishes. This suggests that lower $Hg^{2+}$ ion concentration actually leads to a stronger the Raman fingerprint band, which is reminiscent of the recent "inverse sensitivity" mechanism discussed in Au nanostructures[48]. Fig. 5a shows the representative SERS spectra of the GT-rich oligonucleotide under coordination of various concentrations of $Hg^{2+}$ ranging from 0 to $4\times10^6$ ppb, where the Hoogsteen bands at ~1,485 $cm^{-1}$ are inversely proportional to the $Hg^{2+}$ concentrations. The intensities were normalized and plotted statistically as shown in Fig. 5b where a threshold level is defined as three times of standard deviation from the blank sample is used to identify detection limit of the assay (L.O.D). The bar graph indicates that concentrations of $Hg^{2+}$ ranging from $2\times10^{-4}$ to $4\times10^{6}$ ppb could be detected, and the lowest detectable concentration ($2\times10^{-4}$ ppb) is four orders of magnitude lower than the exposure limit allowed by EPA. Strikingly, the detection limit of the assay far exceeds all reported sensitivity of $Hg^{2+}$ detections, including the L.O.D of 20 ppb[29] or 2 ppb[30] for colorimetric detections using DNA-functionalized gold nanoparticles, 0.2 ppb for DNA-based machine[30] or fluorescence polarization enhanced by gold nanoparticles[31].

Different metallic ions (such as $Ca^{2+}$, $Cu^{2+}$, $Cd^{2+}$, $Mg^{2+}$, $Ni^{2+}$, $Zn^{2+}$) have been used to investigate the selectivity of the logic gate. The results in the Fig. 5c show that these metallic ions do not prevent the formation of Hoogsteen hydrogen bonding, their corresponding normalized Hoogsteen band intensities are much higher than that of the $Hg^{2+}$-treated sample (Fig. 5d). Therefore, the MetaSERS based logic gate has not only ultrahigh sensitivity but also good selectivity for the detection of $Hg^{2+}$.

In summary, we have demonstrated the Vis-NIR tunable alphabetical metamaterials which enable unique ability for optically controlling the hot-spots. The optical response can be readily tuned in Vis-NIR range by tailoring the size and shape of the resonators. We have also demonstrated for the first time the use of metamaterials as SERS-based logic gate operations and for the detection of mercury ions with ultrahigh sensitivity and selectivity based on the specific conformation modulation of a GT-rich oligonucleotide. The most notable attributes of the MetaSERS-based logic gates developed in this effort are their label-free measurement, sensitivity, reversibility, reproducibility, and simplicity. The novel concept of using MetaSERS will open up a new approach for molecular logic gates that can be possibly operated down to single molecule level, and will be beneficial for a variety of applications such as clinical diagnostics, environmental monitoring, food safety analysis, and biological computations.

**Methods**

**Fabrication of alphabetical metamaterials.** The metamaterials with different bar widths from 30 to 80 nm were fabricated on 0.7 mm-thick ITO/glass substrates over an area of 40 μm x 40 μm by electron beam lithography (EBL). Commercial electron beam resist polymethyl methacrylate was spin-coated at 4,000 rpm for 1 min on the ITO/glass, and baked at 180 °C for 20 min. The metamaterials patterns (geometrical parameters and periodicity of the unit cell are described in the Supplementary Information) were produced using a JEOL 7001 F SEM equipped with a nanometer pattern generation system (NPGS), and then developed in 1:3 methyl isobutyl ketone:isopropyl alcohol (MIBK:IPA) developer for 90 s. After the development, 30 nm Au film following a 2-nm Cr as an adhesive layer was deposited using thermal evaporation deposition (Elite Engineering, Singapore) at a base pressure of $3 \times 10^{-7}$ Torr. Finally, the sample was immersed in acetone for at least 3 hr for lift-off, and washed thoroughly with IPA and water.

**Transmission and SERS measurements.** To evaluate the resonance modes of alphabetical metamaterials, the transmission spectra were conducted using a microspectrophotometer (Craic 20) in the range of 400-2,100 nm. The laser tunable SERS spectroscopy was performed in a back scattering geometry using a Jobin-Yvon HR800 Raman system equipped with a liquid nitrogen-cooled charge-coupled detector (CCD). The laser excitation wavelengths are selected from a Ti-Sapphire laser (Coherent). For pattern tunable experiments (Fig. 3), Jobin-Yvon T64000 was used in a back scattering geometry excited by a diode laser (λ = 785 nm). For all SERS experiments, the laser power was kept bellow 1 mW otherwise stated. For the laser tunable SERS spectroscopy (Fig. 2), an important step is to calibrate the wavelength-dependent laser flux and the equipment response. We used a silicon wafer with <111> orientation as a standard sample for the calibration of the laser flux, while we used a standard tungsten halogen light source (HL-2000, Ocean Optics) to calibrate the equipment response.

**Discrete dipole approximation simulation.** The electric-magnetic field of alphabetical metamaterials is simulated by discrete dipole approximation (DDA) method using the DDSCAT program (version 7.0)[34, 35]. 2-nm grids were used for all simulations.

**Operation of MetaSERS-based logic gates**. The GT-rich DNA (10 μM) was first heated to 90 °C for 10 min and then immediately chilled in ice water for 2 hr. For the INHIBIT logic gate, final effective concentrations of the pretreated GT-rich DNA (2 μM) was then added into different solutions: HEPES buffer (input = 0, 0), HEPES buffer plus 1 mM $Hg^{2+}$ (input = 1, 0), HEPES buffer plus 1 mM $Hg^{2+}$ and 20 mM $K^+$ (input = 1, 1), HEPES buffer plus 20 mM $K^+$ (input = 0, 1). For an OR logic gate operation, an additional amount of 50 mM $I^-$ ion was subsequently introduced into the samples. The samples were then incubated for 2 hr at room temperature. Subsequently, an aliquot of the reacted solutions containing the GT-rich DNA and ions was dropped onto the U-shaped SRR substrate. Then, a glass coverslip (thickness no. 1) was placed on the SRR substrate and sealed with parafilm stripes to avoid evaporation. SERS measurement was performed on the SRR substrates using a micro-Raman spectrometer (Horiba-JY T64000) excited with a diode laser ($\lambda$ = 785 nm) in the backscattering configuration. The back scattered signal was collected through a 50 X objective lens, the laser power on the sample surface was measured about 2.5 mW, and acquisition time was 50 s.

For the sensitivity and selectivity experiments, volumes containing final effective concentrations of 2 μM preheated GT-rich DNA, HEPES buffer (50 mM HEPES buffer pH 7.4, 0.1% Triton X-100, 2% dimethyl sulfoxide), and each concentrations of $Hg^{2+}$ ranging from 0 to $4 \times 10^6$ ppb were incubated for 2 hr at room temperature. For evaluating the selectivity, various metallic ions at 1 mM concentration ($Ca^{2+}$, $Cu^{2+}$, $Cd^{2+}$, $Mg^{2+}$, $Ni^{2+}$, $Zn^{2+}$) were used instead of $Hg^{2+}$. The samples were also incubated for 2 hr at room temperature before the SERS analyses as described above.

**Acknowledgments**

The author Q.X. thanks the strong support from Singapore National Research Foundation through a Fellowship grant (NRF-RF2009-06), the support from Singapore Ministry of Education via two Tier 2 grants (MOE2011-T2-2-085 and MOE2011-T2-2-051), and a very strong support from Nanyang Technological University via start-up grant (M58110061). X.X. acknowledges Natural Science Basic Research Plan in Shaanxi Province of China (No. 2012KJXX-27). S.L. acknowledges the support from Singapore Ministry of Education via Tier 2 grants (MOE2011-T2-2-085) and the support from Nanyang Technological University via start-up grant. P.T. acknowledges the support from the special funds for Major State Basic Research of China, Contract No. 2009CB929301, the National Natural Science Foundation of China, Grants No. 11225421 and No. 10934007.

**Author contributions.** J.Z., C.C., and Q.X. conceived the ideas, P.T. designed the wavelength-tunable Raman experiments, C.C., X.X and J.Z. performed the experiments, C.L. and S.L. conducted the DDA simulation, J.Z., C.C., and Q.X. analyzed the data and wrote the manuscript, all the authors commented on the manuscript.

**Competing financial interests.** The authors declare no competing financial interests.

**Figures and Captions**

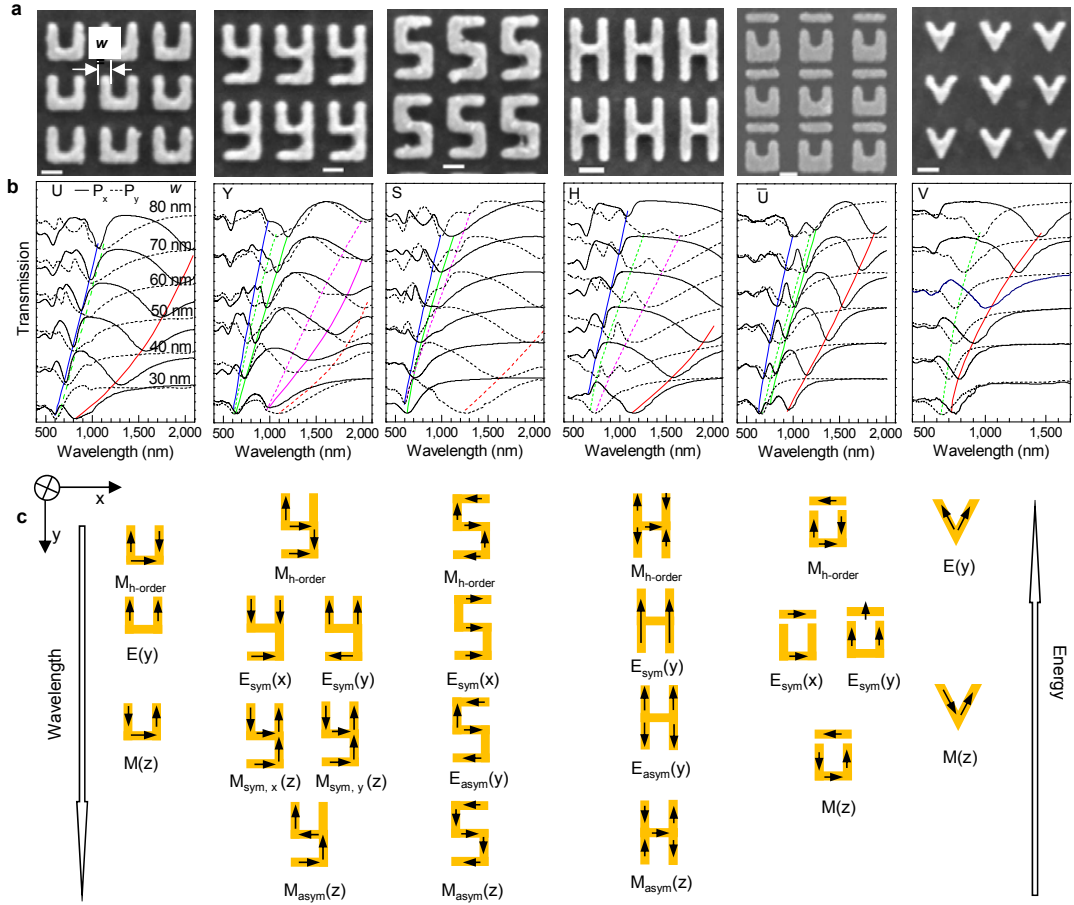

**Figure 1. SEM images, transmission spectra and schematic diagrams of the plasmon hybridization. a,** A typical SEM images of U, Y, S, H, U-bar and V shaped gold metamaterials with $w$ = 40 nm fabricated on a flexible PEN or ITO/Glass substrate. **b,** Transmission spectra of SRRs with different widths of $w$ = 80-30 nm for U, Y, S, H, U-bar and V shapes. The solid curves correspond to the $P_x$ polarization, while the dashed curves correspond to the $P_y$ polarization. The color lines highlight the trends of resonance depending on the size of unit. **c,** Dipole current distribution of the plasmon hybridization modes for different shapes. For each shape, the modes are arranged by the order of wavelength (energy), with two degenerate cased for Y and U-bar shapes. The corresponding simulated results can be found in Figure S2.

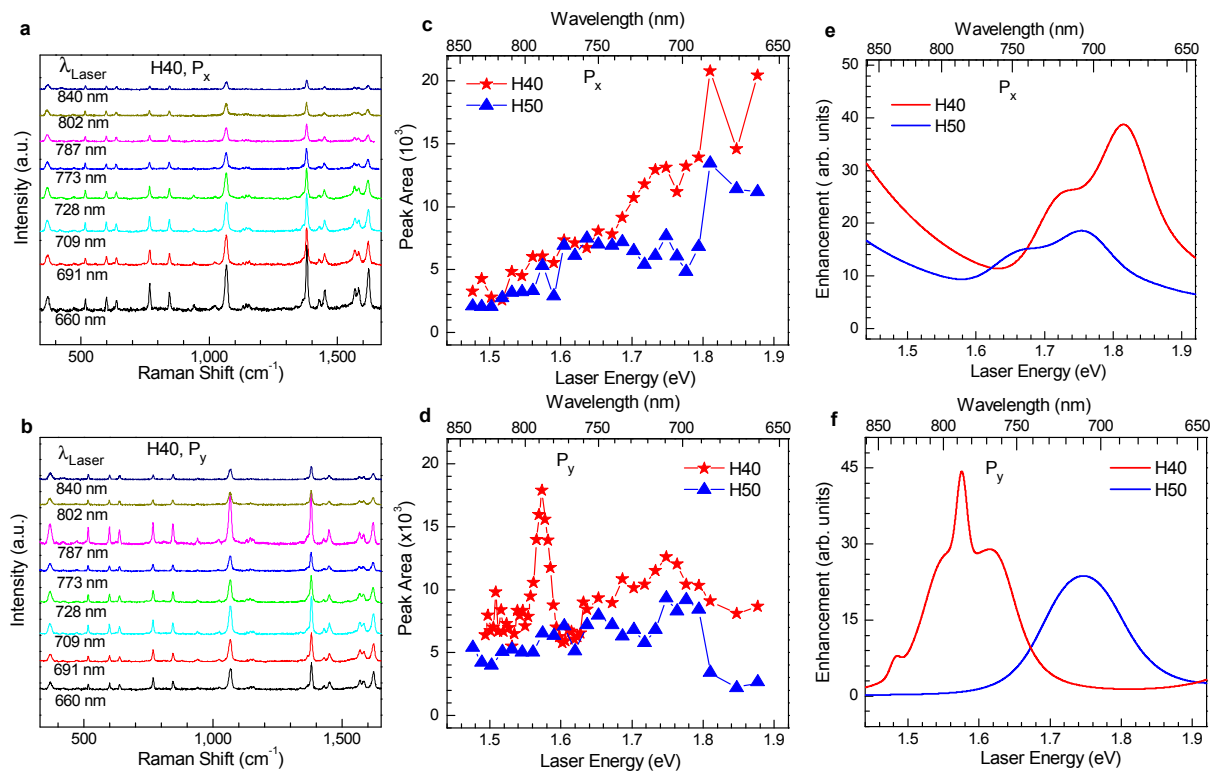

**Figure 2. Excitation wavelength dependence of experimental SERS spectra and simulated enhancement spectra at $P_x$ and $P_y$ polarization. a** and **b**, are the selected measured SERS spectra depends on the laser wavelength for H40 at $P_x$ and $P_y$ polarizations. **c** and **d**, are laser energy dependence of the integrated area intensity of ring-ring stretching mode of benzene functional groups at 1,380 cm$^{-1}$ for H40 and H50 pattern at two polarization configuration. **e** and **f**, are the simulated enhancement of SERS signal corresponding to the **c** and **d**. The enhancement factor value was calculated from the product of enhancement factors at the laser wavelength and that at the scattering wavelength, *i.e.* $EF = E^2(\lambda_{laser}) \times E^2(\lambda_{scattering})$. The scattering peak is 1,380 cm$^{-1}$.

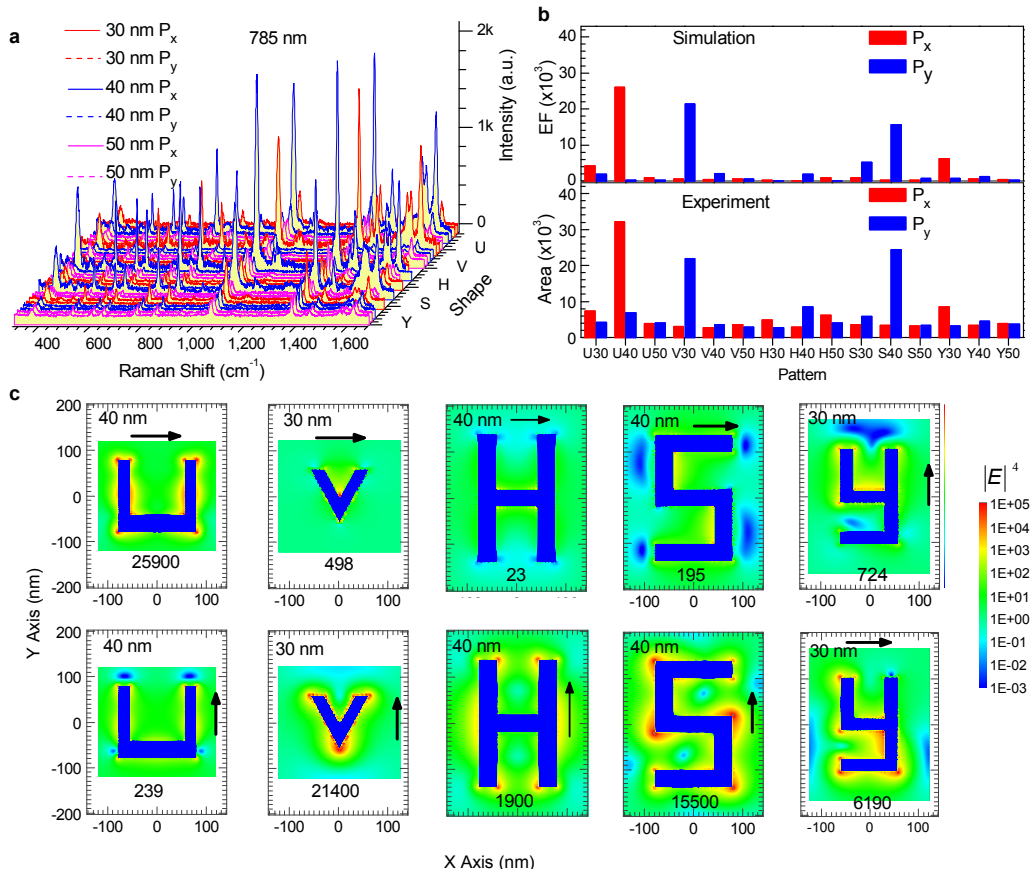

**Figure 3. Shape tunable SERS spectra and simulation results at 785 nm excitation. a,** The experimental SERS spectra depending on the shape and width ($w$ = 30 nm for red line, $w$ = 40 nm for blue line, $w$ = 50 nm for pink line) of the alphabetical metamaterials at 785 nm laser with two cross-polarization configurations. The solid lines and the dash lines represent for the laser polarizations which are parallel and vertical to the gap of the alphabet metamaterials, respectively. **b**, The statistics results of experimental SERS spectra in **a** and simulated average enhancement factor for S, H and Y shapes from $w$ = 30 nm to $w$ = 50 nm with two polarization configurations. **c**, The simulated contour plot of SERS enhancement factor $|E|^4$ distribution for the pattern with the highest enhancement. The black arrows correspond to the laser polarizations. The right corner values are the width *w* of the SRRs unit, and the bottom values are the average enhancement factor.

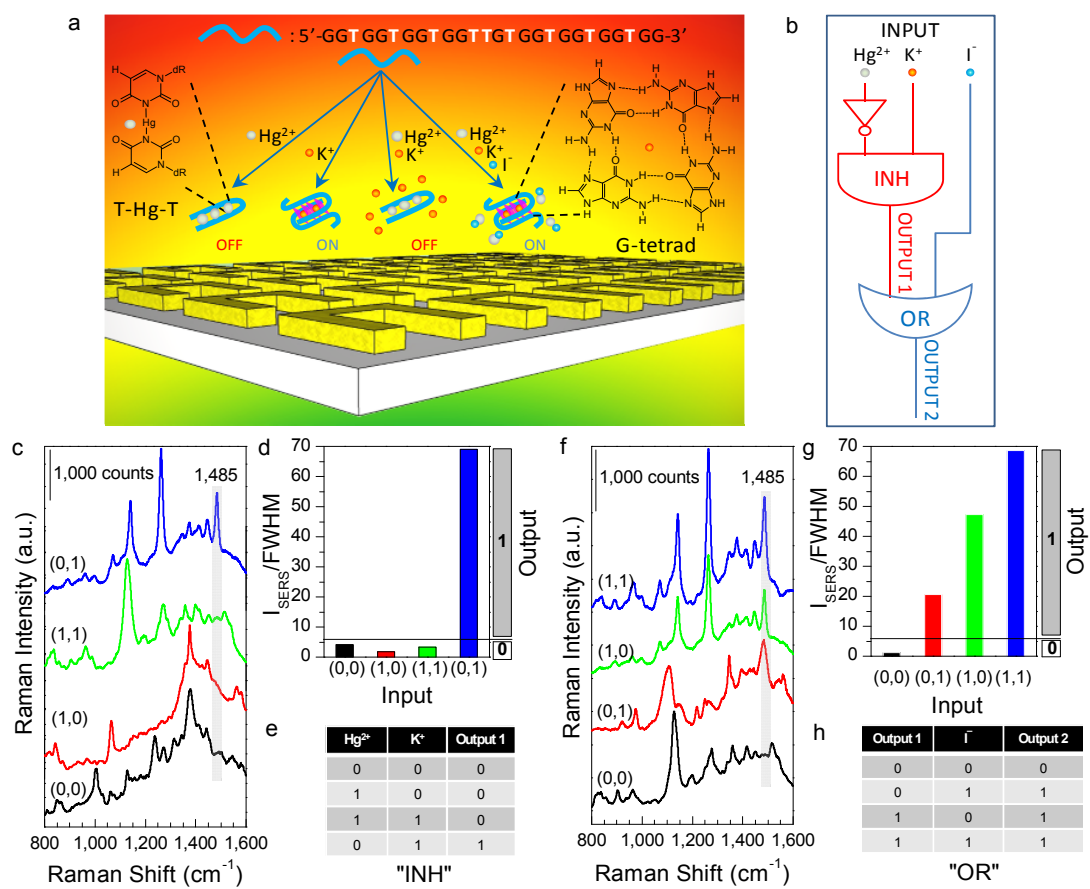

**Figure 4. Combinational logic gate operations (INH, and OR) using MetaSERS**. **a**, Schematic illustration of the Hoogsteen hydrogen bonding activation using the GT-rich oligonucleotide with $Hg^{2+}$, $K^+$, and $I^-$ ions as inputs. The output is translated into "ON" or "OFF" states of SERS signal at ~1,485 $cm^{-1}$, which is a diagnostic marker of the C8=N7-H2 Hoogsteen hydrogen bonding of the folded G-quadruplex structure. **b**, Network map schematically represents the combination of INHIBIT and OR gates for the switchable logic operations. **c**,

SERS spectroscopy using U45 SRRs surface for the monitoring of Hoogsteen band formation at 1,485 cm$^{-1}$ of the GT-rich DNA (2 μM) under coordinating effects of 1 mM Hg$^{2+}$ and 20 mM K$^{+}$. **d**, and **e**, are normalized Raman intensity changes ($I_{SERS}$/FWHM) at 1,485 cm$^{-1}$ and a truth table for the INHIBIT logic gate, respectively. **f**, SERS spectra show that formation of the Hoogsteen band is switchable upon the introduction of 50 mM iodide anion (I$^{-}$). The normalized Raman intensities at 1,485 cm$^{-1}$ are plotted in **g**), and **h**) is a truth table for the OR logic gate operation.

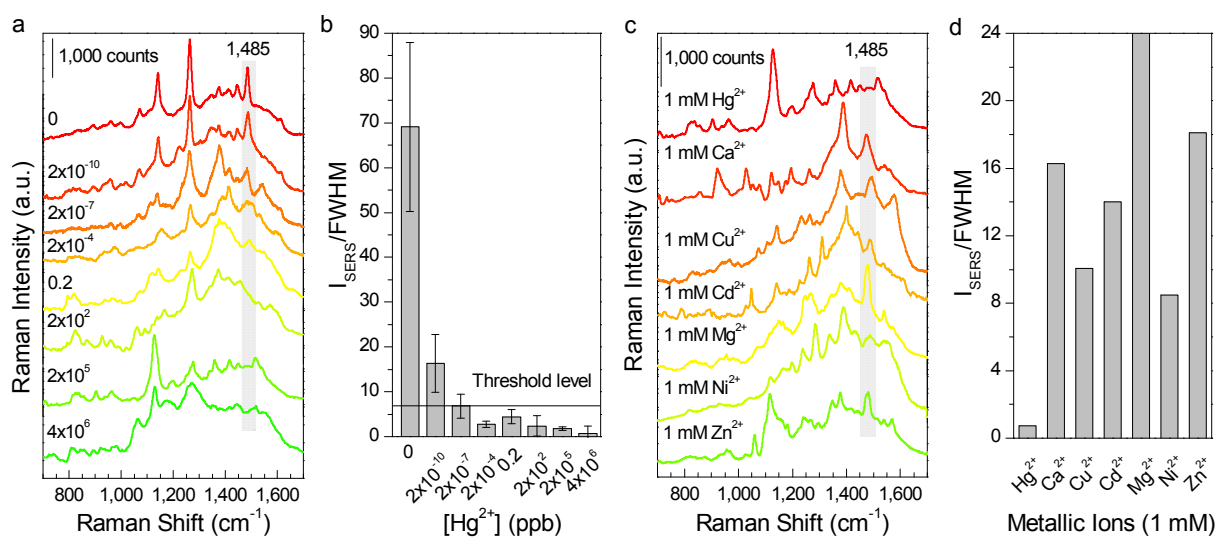

**Figure 5. MetaSERS for the detection of mercury ions. a**, SERS spectra of the GT-rich oligonucleotide under coordination of various concentrations of Hg$^{2+}$ ranging from 0 to 4×10$^{6}$ ppb. The diagnostic Hoosteen band intensities at ~1,485 cm$^{-1}$ show an inverse relationship with the Hg$^{2+}$ concentrations. **b**, Statistic data represent the correlation of normalized SERS intensities at ~1,485 cm$^{-1}$ as a function of Hg$^{2+}$ concentrations for three parallel acquisitions. **c**, SERS spectra of the GT-rich oligonucleotide treated with various metallic ions at 1 mM concentration. **d**, The normalized SERS intensities at 1,485 cm$^{-1}$ indicate that the mercury ions were clearly differentiated from other cations at the same concentration.

# Supplementary Information for

# Tailoring Plasmonic Metamaterials for DNA Molecular Logic Gates

Jun Zhang[1,#], Cuong Cao[1,#], Xinlong Xu[2], Chihao Liow[3], Shuzhou Li[3], Ping-Heng Tan[4], and Qihua Xiong[1,5,*]

[1]Division of Physics and Applied Physics, School of Physical and Mathematical Sciences, Nanyang Technological University, Singapore 637371

[2]Nanobiophotonic Center, State Key Lab Incubation Base of Photoelectric Technology and Functional Materials, Institute of Photonics & Photon-Technology, Northwest University, Xi'an 710069, China

[3]Division of Materials Science, School of Materials Science and Engineering, Nanyang Technological University, Singapore 639798

[4]State Key Laboratory of Superlattices and Microstructures, Institute of Semiconductors, Chinese Academy of Sciences, Beijing 100083, China

[5]NOVITAS, Nanoelectronics Centre of Excellence, School of Electrical and Electronic Engineering, Nanyang Technological University, Singapore, 639798

[*]To whom correspondence should be addressed. Email address: Qihua@ntu.edu.sg.

[#]These authors equally contributed to this work.

## 1. Definition of alphabetical metamaterials

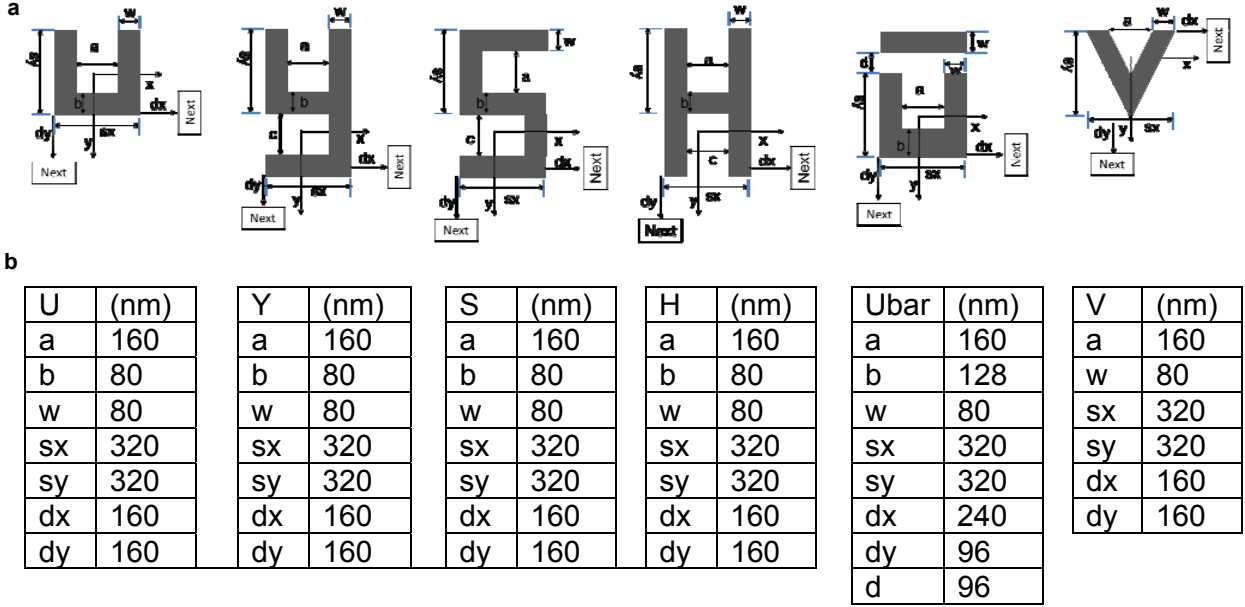

**Figure S1. Geometry definition of alphabetical metamaterials.** **a**, the layout definition of the different alphabet metamaterials with bar width of $w = 80$ nm. **b**, the actual value of every parameter as labeled in **a**. The whole size and periodicity of the unit cell are shrunk accordingly from $w = 80$ nm to $w = 30$ nm in order to tune the resonance peak in Vis-NIR range.

## 2. Mode Identifications: S, H, U-bar, and V Shaped Resonators

In the main text part, we have discussed the mode identifications for the fundamental resonator of U-shape and coupled resonator of Y-shape. Once we understand the physical pictures of those two resonators, we can readily address the mode identifications for other four resonators. The S and H shaped resonators can be considered as two 180°-rotated U-shaped resonator configurations connected side-by-side and back-to-back, respectively. The electric (magnetic) dipole-dipole coupling can also form new coupled electric (magnetic) modes. In the S shape, we observed four resonance modes as highlighted by solid blue, dash green, dash pink and solid red lines in Fig. 1b shown in the main text, which are respectively identified as higher-order magnetic resonance mode $M_{h\text{-order}}$, symmetrically coupled electric modes $E_{sym}(x)$ and asymmetrically coupled electric modes $E_{asym}(x)$, and asymmetrically coupled magnetic mode $M_{asym}(z)$ as shown in Fig. 1c. The $E_{asym}(y)$ mode is very weak and even unresolved when the bar-width is smaller than 50 nm, however it becomes pronounced for larger size resonators (highlighted by pink dashed line). In H shape, besides $M_{h\text{-order}}$, $E_{sym}(y)$ and $M_{asym}(z)$ modes, we also observed an asymmetrically coupled electric mode of $E_{asym}(y)$ in the $P_y$ polarization excitation, which is completely dark and decoupled from the normal incident light if the metamolecule exhibits spatial inversion/reflection symmetry, such as the U shape, in the plan of structure[1]. Here, we propose that the observation of $E_{asym}(y)$ in the H-shaped metamaterial is due to the coupling effects between two U shapes. As there is no phase retardation between the two 180°-rotated U pairs, only a single asymmetric magnetic resonance can be observed in the S and H shapes[2]. The U-bar structure is also known as asymmetric split ring resonators (ASRRs)[3]. It consists of a SRR and a bar. The higher order magnetic resonance $M_{h\text{-order}}$ shows asymmetric alignment of electric dipoles in both of two vertical arms of SRR, and in both of the bar and the bottom arm of SRR. The electric resonance mode is contributed from two degenerated modes of $E_{sym}(x)$ and $E_{sym}(y)$ with parallel alignment of electric-dipoles along $x$- and $y$-direction, respectively. Circulating currents induced by $P_x$-polarized incident light lead to the magnetic resonance $M(z)$. For the V shape, two resonance modes come from the coupling of two dipoles in the angled arms. The $P_y$-polarization of light induces symmetric aligned electric diploes in two arms, resulting in an electric mode $E(y)$. Similar with asymmetric coupling in two non-contacted nanowires[2], the displacement current of asymmetric coupled dipoles between two arms of the V shape also has partial circulating features along the V shape, leading to a resonant excitation of magnetic dipole moment $M(z)$.

3. **DDA Simulation**

Our experiments are supported by simulations of the local electromagnetic fields by discrete dipole approximation (DDA) method in the DDSCAT program[4, 5]. 2-nm grids were used for all simulations. The average and maximum electric field intensities over the nanoparticle surface were calculated for isolate metamolecules in a homogenous dielectric environment where the substrate effect was included explicitly by effective media theory. The data was used to produce contour plots of the intensity on and around the nanoparticle to visualize the location of the hotspots. The extinction efficiency was simulated for multiple wavelengths to produce resonance profile and to determine the resonance modes of the metamolecules. The dipole-current diagram of every resonance mode also was produced from the simulation.

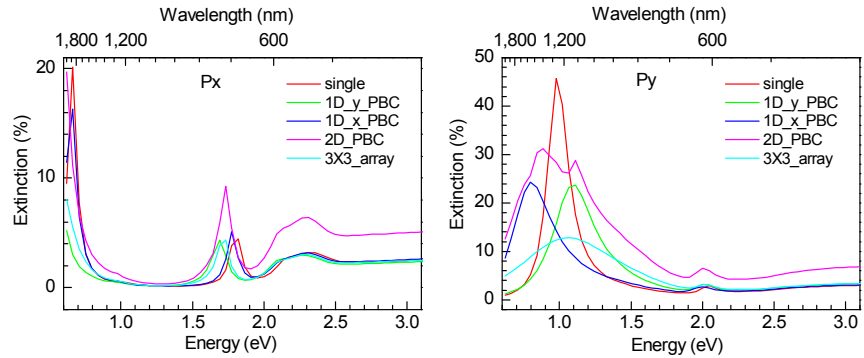

**Figure S2. Simulated extinction spectra depending on the dimensions of selected metamolecule units.** The demonstrated metamaterials is the H shape with bar-width $w$ = 30 nm. The Px and Py are the polarization directions of incident light along to the $x$- and $y$-directions, respectively. The red line is simulated by single metamolecule unit cell. The green and blue lines are simulated by the one dimensional

periodical boundary condition along to *y*- and *x*-directions, respectively. Pink line is simulated by two dimensional periodical boundary conditions. Cyan line is simulated by three-by-three array unit cells.

When the space between metamolecules is very close, the coupling of electric-magnetic fields between neighboring metamolecules cannot be ignored anymore and the suitable periodical boundary conditions (PBC) must be considered during the simulation. In order to examine how extinction spectra depend on the PBC of choice, we use the H shape metamaterials of *w* = 30 nm with the smallest space as an example to test it. The simulated results of two $P_x$ and $P_y$ polarization configurations are shown in Figure S2. For the $P_x$ polarization, both of the mode number and spectral shape are almost identical to each other, and the resonant peaks are relatively shifted for different PBC of choice. For the $P_y$ polarization, except the peak shift similar as $P_x$ polarization condition, the spectra widths and profiles show a much stronger PBC dependence than the $P_x$ case. As the increasing of PBC dimension from single metamolecule to 3×3 arrays, the width of spectra, especially the low-energy modes, is broadening. The two dimensional PBC simulation shows a multi-mode profile of low-energy mode. This multi-mode profile is caused by the anisotropy of *x*-direction PBC and *y*-direction PBC due to the antisymmetry of the H shape. When PBC increases to the 3×3 arrays, this anisotropy reduces and spectrum shows a broad single peak. Qualitatively, the single metamolecule approximation can give a very good simulation for the real resonance spectra of alphabetical metamaterials with a bar-width of $w \geq 30$ nm. Therefore, in our DDA simulations, we take a single metamolecule unit to produce the data of hot-spot contour plot in Figure 3 and resonance mode identifications in Figure S3. In order to give more quantitative results, we used the 2D PBC to produce the resonant profile of SERS signal in Figure 2e and 2f.

Besides the analysis based on *plasmon hybridization* or *dipole-dipole coupling theory,* we also use DDA simulation to reconfirm the resonant profile and resonant mode identification for all alphabetical metamolecules as shown in Figure S3. The dipole-current distributions corresponds the Figure 1c in the main text part. We import the SEM image of metamaterials as the simulated structure in order to obtain more accurate results. We found the simulation by using 60 nm thicknesses is better agreed with the experimental data than the simulation using 30 nm. The physical mechanism is the deference of effective permittivity between evaporated gold film in our resonator and crystalline gold film in simulation, which has been discussed in Linden *et al.*[6]. In evaporated gold film in experimental metamaterials, the electron should be experienced more scattering than in crystalline gold, as a results the experimental resonance modes will blue shift comparing the simulated results. In our simulation, we use the permittivity value of crystalline gold film to simulate our evaporated non-crystalline gold film. Consequently, the much thicker of the gold film we simulated, the permitivity is much closer to the value of the crystalline gold film and thus the results are more agreed with the experimental data. If we define the accuracy as $\frac{|\lambda_{exp} - \lambda_{sim}|}{\lambda_{exp} + \lambda_{sim}}$, the deviation is found to be within 10%.

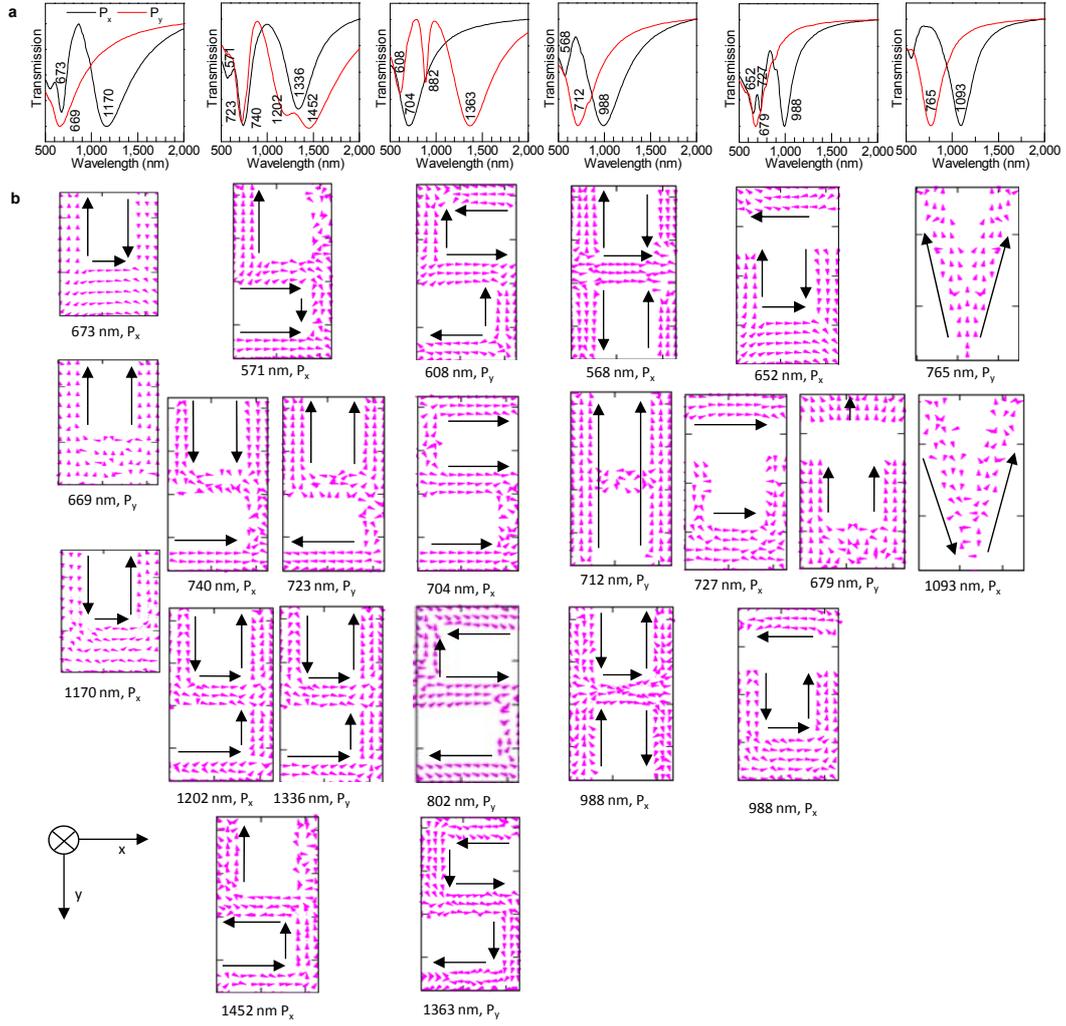

**Figure S3. DDA simulation of transmission spectra and corresponding dipole current distributions**. **a**, the simulated transmission spectra of U, Y, S, H, U-bar and V shapes (from left to right) with a bar-width $w$ = 30 nm and a thickness of 60 nm. The 60 nm thickness was chosen in order to make the simulated gold film attributable to more scattering to match the real scattering in evaporated non-crystalline gold film. **b**, the dipole current distributions of different resonant peaks in **a**. the $P_x$ and $P_y$ are the polarizations of incident light along x- and y-directions, respectively.

## 4. Enhancement factor calculation in DDA simulation and experimental estimation

According to the Mie scattering theory, the electromagnetic enhancement factor (EF) is the production of incident light and scattering light enhancement, *i.e.* $EF_{EM} = \frac{|E(\lambda_{laser})|^2 \times |E(\lambda_{scatt})|^2}{|E_0(\lambda_{laser})|^2 \times |E_0(\lambda_{scatt})|^2}$, where the $|E(\lambda_{laser})|^2$ and $|E(\lambda_{scatt})|^2$ (or $|E_0(\lambda_{laser})|^2$ and $|E_0(\lambda_{scatt})|^2$) correspond the intensity of localized (or normal) electromagnetic field at incident laser and scattering light wavelength, respectively. During enhancement calculation, usually both of the $|E_0(\lambda_{laser})|^2$ and $|E_0(\lambda_{scatt})|^2$ were normalized to unity. Hence, the enhancement factor can be written as $EF_{EM} = |E(\lambda_{laser})|^2 \times |E(\lambda_{scatt})|^2$. Because the dispersion relation of $|E(\lambda_{laser})|^2$ is proportional to the extinction spectra, we use the calculated extinction spectra to calculate the SERS enhancement depending on the excitation wavelength in Figure 2e and 2f. Considering that the scattering wavelength $\lambda_{scatt}$ is very close to the excitation wavelength $\lambda_{laser}$ in Raman spectroscopy, the enhancement factor can be further simplified as zero Stokes shift production of $EF_{EM} = |E(\lambda_{laser})|^4$. This formula was used to calculate the hot-spot distribution and enhancement factor in Figure 3b and 3c.

In practical experiments, the enhancement factor (EF) of SERS spectra is defined as follows:

$$EF = \frac{I_{sers}/N_{sers}}{I_{norm}/N_{norm}}, \tag{S-1}$$

where the $I_{sers}$ and $I_{norm}$ are the integral intensities of SERS spectra and normal Raman spectra, respectively. The $N_{sers}$ and $N_{norm}$ are the numbers of molecules contributed to the SERS signal and normal Raman signal, respectively.

For the normal Raman spectra (no SERS-active substrate), we used 2-Napthalenthiol (2-NAT) powder sample as a reference. The number of molecules contributed to normal Raman spectra can be calculated from the equation as follows:

$$N_{norm} = \pi r_{spot}^2 h \times \rho_{2-NAT} \times N_A / M_{2-NAT}, \tag{S-2}$$

where the $\pi r_{spot}^2$ is the area of laser spot at the sample, $h$ is the laser penetration depth in 2-NAT powder sample, which equal to the thickness of 2-NAT powder ($h$ = 1 mm) due to total transparent of 2-NAT for 785 nm laser[7], $\rho_{2-NAT} = 1.176 g/cm^3$ is the density of 2-NAT powder, $N_A$ is the Avogadro constant, and $M_{2-NAT}$ = 160.24 g/mol is the molecular weight of 2-NAT powder. By substituting those values into the equation (S-2), the new expression of molecule number involving in the SERS signal are as follows:

$$N_{norm} = 4.42 nm^{-3} \times \pi r_{spot}^2 h, \tag{S-3}$$

For the SERS spectra of monolayer 2-NAT covered on alphabetical metamaterials, the molecule number contributing to the SERS signal can be calculated by the following equation:

$$N_{SERS} = \frac{(\pi r_{spot}^2 \cdot R_{lattice}) \frac{S_{hot-spot}}{S_{lattice}}}{S_{2-NAT}}, \quad (S-4)$$

where the fill factor $R_{lattice}$ is a ratio of the surface area of gold metamaterials to the area of whole unit cell, $S_{hot-spot}$ is the hot-spot area, $S_{lattice}$ is the gold metamaterials area, and $S_{2-NAT} = 0.42$ nm$^2$ is the area of single 2-NAT molecule[7], which is also called *molecule boot-print*. If we define the ratio of hot-spot area to gold metamaterials area as $R_{hot-spot} = \frac{S_{hot-spot}}{S_{lattice}}$ and combine the equation (S-1), (S-3) and (S-4), the EF can be written as follows:

$$EF = 1.856 nm^{-1} h \frac{1}{R_{lattice} R_{hot-spots}} \frac{I_{sers}}{I_{norm}}. \quad (S-5)$$

For a given pattern with certain size, the $R_{lattice}$ can be easily calculated from the pattern definition as shown in Figure S1. As an average evaluation, we suppose that the hot-spot area is equal to the gold pattern area, *i.e.* $R_{hot-spots} = 1$, which is an upper limit of $R_{hot-spots}$ because the hot-spot area is always less than the gold pattern area as shown in Figure 3c.

In order to check the enhancement factor of the alphabet metamaterials, we measured the SERS spectra of monolayer 2-NAT molecules on the metamaterials and power 2-NAT with 1 mm thickness with the same experimental conditions. The results are shown in the Table S-1. In order to show the statistical results of enhancement factors for different patterns, we plot the measured enhancement factors versus numbers of samples as shown in Figure S4. We found that most of the samples contribute to an EF of ~10$^7$, albeit even there are two samples contributing to an EF ~10$^8$.

**Table S-1. The enhancement factor calculated for different structures at 785 nm excitation.** The normal Raman spectra were measured from a 2-NAT powder with thickness of 1 mm. The $R_{hot-spots}=1$ was taken as an average enhancement evaluation.

| Pattern | h (1mm) | $P_x$(785nm, 1380 cm$^{-1}$) | | | $P_y$(785nm, 1380 cm$^{-1}$) | | |
|---|---|---|---|---|---|---|---|
| | | $I_{sers}/I_{bulk}$ | $R_{lattice}$ | EF(10$^7$) | $I_{sers}/I_{bulk}$ | $R_{lattice}$ | EF(10$^7$) |
| S30 | 1 | 7.08 | 0.53 | 2.48 | 12.92 | 0.53 | 4.53 |
| S40 | 1 | 1.88 | 0.53 | 0.66 | 20.89 | 0.32 | 12.14 |
| S50 | 1 | 3.10 | 0.53 | 1.09 | 3.05 | 0.53 | 1.07 |
| H30 | 1 | 9.28 | 0.53 | 3.26 | 4.51 | 0.53 | 1.58 |
| H40 | 1 | 10.31 | 0.53 | 3.62 | 19.21 | 0.53 | 6.76 |
| H50 | 1 | 8.06 | 0.53 | 2.83 | 2.60 | 0.53 | 0.91 |
| V30 | 1 | 3.32 | 0.32 | 1.93 | 18.7 | 0.32 | 10.56 |
| V40 | 1 | 2.64 | 0.32 | 1.53 | 3.38 | 0.53 | 1.19 |
| V50 | 1 | 4.11 | 0.32 | 2.39 | 4.45 | 0.53 | 1.56 |
| Y30 | 1 | 12.37 | 0.32 | 7.19 | 0.37 | 0.53 | 0.13 |
| Y40 | 1 | 4.14 | 0.53 | 1.45 | 14.02 | 0.53 | 4.92 |
| Y50 | 1 | 5.61 | 0.53 | 1.97 | 5.87 | 0.53 | 2.06 |

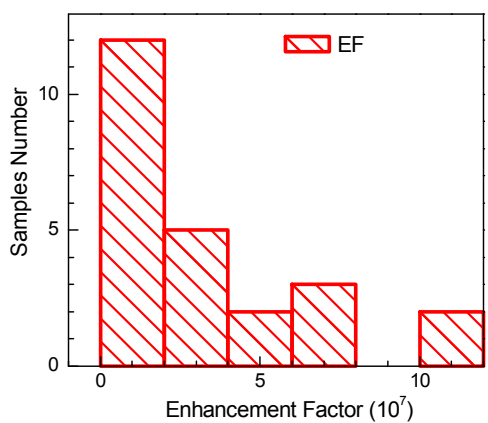

**Figure S4. The histogram plot of SERS enhancement factors.**

## 5. Materials

The GT-rich oligonucleotide DNA (5'-GGT GGT GGT GGT TGT GGT GGT GGT GG-3') was purchased from Integrated DNA Technologies, Singapore. HEPES buffer (50 mM HEPES buffer pH 7.4, 0.1% Triton X-100, 2% dimethyl sulfoxide), $Hg(ClO_4)_2 \cdot H_2O$, and other essential chemicals were of analytical grade and obtained from Sigma-Aldrich, Singapore unless otherwise stated. All experiments were done using DNA-free water (1st Base, Singapore).